\documentclass[12pt]{iopart}

\usepackage{graphicx}
\usepackage{amssymb}
\usepackage{iopams}
\usepackage{amsthm}
\usepackage{latexsym}
\usepackage{pst-all}
\usepackage{capt-of}
\usepackage{epsfig}

\newcommand{\E}{{\cal{E}}}

\newcommand{\I}{{\rm i}}
\renewcommand{\a}{\alpha}
\renewcommand{\d}{{\rm d}}

\newcommand{\be}{\begin{equation}}
\newcommand{\ee}{\end{equation}}
\newcommand{\bea}{\begin{eqnarray}}
\newcommand{\eea}{\end{eqnarray}}

\def\J#1#2#3#4{#1 {\it #2} {\bf #3} #4}
\def\PTP{\it Prog. Theor. Phys.}
\def\PRL{\it Phys. Rev. Lett.}
\def\PRD{{\it Phys. Rev.} D}
\def\PR{\it Phys. Rev.}

\def\APNY{\it Ann. Phys. (NY)}

\def\JMP{\it J. Math. Phys.}
\def\CQG{\it Class. Quantum Grav.}

\def\PLA{\it Phys. Lett. A}

\begin{document}

\title[On the properties of EMR equatorially antisymmetric solutions]
{On the properties of the Ernst--Manko--Ruiz equatorially
antisymmetric solutions}

\author{Jordi Sod--Hoffs$^1$\footnote[1]{e-mail:
jordi@fis.cinvestav.mx} and Egor D Rodchenko$^2$}

\address{Departamento de F\'{i}sica, Centro
de Investigaci\'{o}n y de Estudios Avanzados del IPN, A.P. 14-740,
07000 M\'{e}xico D.F., Mexico}
\address{$^2$ Department of Quantum Statistics and Field Theory,
Lomonosov Moscow State University, Moscow 119899, Russia}

\begin{abstract}
Two new equatorially antisymmetric solutions recently published by
Ernst {\it et al} are studied. For both solutions the full set of
metric functions is derived in explicit analytic form and the
behavior of the solutions on the symmetry axis is analyzed. It is
shown in particular that two counter--rotating equal
Kerr--Newman--NUT objects will be in equilibrium when the
condition $m^2+\nu^2=q^2+b^2$ is verified, whereas two
counter--rotating equal masses endowed with arbitrary magnetic and
electric dipole moments cannot reach equilibrium under any choice
of the parameters, so that a massless strut between them will
always be present.
\end{abstract}


\section{Introduction}

In the recent papers \cite{EMR1,EMR2} the notion of equatorial
antisymmetry has been introduced for stationary axisymmetric
electrovac spacetimes and two new exact solutions of that type
have been constructed within the framework of Sibgatullin's method
\cite{Sib,MSi}. Since Ernst {\it et al} presented their solutions
only in terms of the Ernst complex potentials $\E$ and $\Phi$
\cite{Ern}, it would be of interest to have the complete metrics
related to those solutions because they could facilitate the
analysis of physical properties of the new equatorially
antisymmetric spacetimes. Another aspect of the Ernst--Manko--Ruiz
(EMR) solutions requiring the knowledge of the respective metric
fields is the following: these solutions describe some binary
systems of two counter--rotating masses in the presence of the
electromagnetic field, so with the aid of the analytical
expressions for the metric functions it would be possible to
consider the equilibrium problems of two counter--rotating
constituents in the EMR spacetimes.

Bearing in mind the above two main objectives, in section~2 of the
present paper the metrical fields of the EMR solutions will be
derived using expansions of the determinantal formulae of paper
\cite{RMM} in the $N=2$ case. Furthermore, in section~3 the
obtained analytical formulae fully defining the EMR spacetimes
will be utilized for the resolution of the equilibrium problem of
two equal counter--rotating Kerr--Newman--NUT particles. Apart
from the behavior of EMR solutions on the symmetry axis, the
stationary limit surfaces, magnetic lines of force, ring
singularities and some limits of these solutions will be also
considered.

\section{The Ernst potentials and metric functions of EMR solutions}

The equatorially antisymmetric EMR solutions belong to the $N=2$
subclass of the analytically extended multisoliton solution whose
Ernst potentials $\E$ and $\Phi$ in the general $N=2$ case are
given by the formulae \cite{RMM} \bea \fl {\cal E}=E_+/E_-,
\qquad \Phi=F/E_-, \nonumber\\
\fl E_\pm=\left|\begin{array}{cccc} 1 & 1 & \ldots & 1
\vspace{0.15cm}\\ \pm 1 & {\displaystyle \frac{r_1}{\a_1-\beta_1}}
& \ldots & {\displaystyle \frac{r_{4}}{\a_{4}-\beta_1}}
\vspace{0.25cm} \\ \pm 1 & {\displaystyle
\frac{r_1}{\a_1-\beta_2}} & \ldots & {\displaystyle
\frac{r_{4}}{\a_{4}-\beta_2}} \vspace{0.25cm} \\
0 & {\displaystyle \frac{h_1(\a_1)}{\a_1-\bar\beta_1}} & \ldots &
{\displaystyle \frac{h_1(\a_{4})}{\a_{4}-\bar\beta_1}}
\vspace{0.25cm}\\ 0 & {\displaystyle
\frac{h_2(\a_1)}{\a_1-\bar\beta_2}} & \ldots & {\displaystyle
\frac{h_2(\a_{4})}{\a_{4}-\bar\beta_2}}\\
\end{array}\right|, \quad
F=\left|\begin{array}{cccc} 0 & f(\a_1) & \ldots & f(\a_{4})
\vspace{0.15cm}\\ -1 & {\displaystyle \frac{r_1}{\a_1-\beta_1}} &
\ldots & {\displaystyle \frac{r_{4}}{\a_{4}-\beta_1}}
\vspace{0.25cm}\\ -1 & {\displaystyle \frac{r_1}{\a_1-\beta_2}} &
\ldots & {\displaystyle \frac{r_{4}}{\a_{2N}-\beta_2}}
\vspace{0.25cm}\\ 0 & {\displaystyle
\frac{h_1(\a_1)}{\a_1-\bar\beta_1}} & \ldots & {\displaystyle
\frac{h_1(\a_{4})}{\a_{4}-\bar\beta_1}} \vspace{0.25cm}\\ 0 &
{\displaystyle \frac{h_2(\a_1)}{\a_1-\bar\beta_2}} & \ldots &
{\displaystyle
\frac{h_2(\a_{4})}{\a_{4}-\bar\beta_2}}\\
\end{array}\right|, \nonumber\\ \fl r_n=\sqrt{\rho^2+(z-\a_n)^2},
\label{EF_gen} \eea  where $\beta_l$ are arbitrary complex
parameters, $\a_n$ can take on arbitrary real values or occur in
complex conjugate pairs, and a bar over a symbol means complex
conjugation; $h_l(\a_n)$ and $f(\a_n)$ are constant objects
defined as \be h_l(\a_n)=\bar e_l+2\bar f_l f(\a_n), \quad
f(\a_n)=f(z)|_{z=\a_n}, \label{hln} \ee $e_l$ and $f_l$ being
coefficients in the expressions of the potentials $\E$ and $\Phi$
on the symmetry axis: \be e(z)=1+\frac{e_1}{z-\beta_1}
+\frac{e_2}{z-\beta_2}, \quad f(z)=\frac{f_1}{z-\beta_1}
+\frac{f_2}{z-\beta_2}. \label{axis_gen} \ee

The first of the EMR solutions (henceforth referred to as
solution~I), representing two counter--rotating electrically and
magnetically charged masses, is defined by the axis data \bea e(z)
= \frac{z-k-m-i(a+\nu)}{z-k+m-i(a-\nu)} \cdot
\frac{z+k-m+i(a-\nu)}{z+k+m+i(a+\nu)}, \nonumber\\
f(z) = \frac{2(q+i b)z}{[z-k+m-i(a-\nu)][z+k+m+i(a+\nu)]},
\label{axis1} \eea where the parameters $m$, $a$, $q$, $b$ are the
mass, angular momentum per unit mass, electric charge and magnetic
charge, respectively, $\nu$ is the NUT parameter and $k$ is half
the coordinate distance between the masses. Using (\ref{axis1})
one easily obtains the quantities \bea \fl
e_1=-\frac{2(m+\I\nu)[k-m+\I(a-\nu)]}{k+\I a}, \quad
e_2=-\frac{2(m+\I\nu)[k+m+\I(a+\nu)]}{k+\I a}, \nonumber\\
\fl \beta_1=k-m+\I(a-\nu), \quad \beta_2=-k-m-\I(a+\nu),
\nonumber\\
\fl f_1=\frac{(q+\I b)[k-m+\I(a-\nu)]}{k+\I a}, \quad
f_2=\frac{(q+\I b)[k+m+\I(a+\nu)]}{k+\I a}, \label{ebf1} \eea by
decomposing $e(z)$ and $f(z)$ into simple fractions, and also the
corresponding parameters $\a_n$, namely, \bea \a_1&=&\a_+, \quad
\a_2=\a_-, \quad
\a_3=-\a_-, \quad \a_4=-\a_+, \nonumber\\
\a_\pm&=&\sqrt{\delta\pm2d}, \quad
\delta=m^2+k^2+3\nu^2-a^2-2(q^2+b^2), \nonumber\\
d&=&[(m^2+\nu^2-q^2-b^2)(k^2+2\nu^2-a^2-q^2-b^2)-(m\nu-ka)^2]^{1/2},
\label{alfas1} \eea as roots of the algebraic equation \be
e(z)+\bar e(z)+2f(z)\bar f(z)=0. \label{eq_alg} \ee

Formulae (\ref{ebf1}) and (\ref{alfas1}) fully determine the
respective quantities $f(\a_n)$ and $h_l(\a_n)$ that appear in the
determinants (\ref{EF_gen}). The potentials $\E$ and $\Phi$
calculated in \cite{EMR2} for the data (\ref{axis1}) have the form
\bea {\cal E}=\frac{A-B}{A+B},
\quad \Phi=\frac{C}{A+B}, \nonumber\\
\fl A=(m^2+\nu^2-q^2-b^2)\{[(m^2+k^2+\nu^2+a^2)^2-4(mk+a\nu)^2] \nonumber\\
\times(R_+-R_-)(r_+-r_-) -
\delta\a_+\a_-(R_++R_-)(r_++r_-)\} \nonumber\\
+2\a_+\a_-\{[(m^2+\nu^2-q^2-b^2)(m^2-k^2+a^2-\nu^2) +2(m\nu-ka)^2]
\nonumber\\ \times(R_+R_-+r_+r_-)
+2id(m\nu-ka)(R_+R_--r_+r_-)\}, \nonumber\\
\fl B=4d\a_+\a_-(m+i\nu)\{[m^2+\nu^2-q^2-b^2+i(m\nu-ka)] \nonumber\\
\times(R_++R_--r_+-r_-)-d(R_++R_-+r_++r_-)\}, \nonumber\\
\fl C=4d\a_+\a_-(q+ib)\{[m^2+\nu^2-q^2-b^2+i(m\nu-ka)] \nonumber\\
\times(R_++R_--r_+-r_-)-d(R_++R_-+r_++r_-)\}, \nonumber\\
R_\pm=\sqrt{\rho^2+(z\pm\a_+)^2}, \quad
r_\pm=\sqrt{\rho^2+(z\pm\a_-)^2}. \label{EF1} \eea

The other EMR equatorially antisymmetric solution (henceforth
solution~II), describing a pair of counter--rotating masses
endowed with electric and magnetic dipole moments, arises from the
axis data \begin{eqnarray} e_{+}(z) =
\frac{z-k-m-\I(a+\nu)}{z-k+m-\I(a-\nu)} \cdot
\frac{z+k-m+\I(a-\nu)}{z+k+m+\I(a+\nu)}, \nonumber\\
f_{+}(z) = \frac{2(\chi+\I
c)}{[z-k+m-\I(a-\nu)][z+k+m+\I(a+\nu)]}, \label{axis2}
\end{eqnarray} and the corresponding quantities \bea
\fl e_1=-\frac{2(m+\I\nu)[k-m+\I(a-\nu)]}{k+\I a}, \quad
e_2=-\frac{2(m+\I\nu)[k+m+\I(a+\nu)]}{k+\I a}, \nonumber\\
\fl \beta_1=k-m+\I(a-\nu), \quad \beta_2=-k-m-\I(a+\nu),
\nonumber\\
\fl f_1=\frac{\chi+\I c}{k+\I a}, \quad f_2=-\frac{\chi+\I c}{k+\I
a}, \label{ebf2} \eea together with the parameters \bea
\a_1&=&\a_+, \quad \a_2=\a_-, \quad
\a_3=-\a_-, \quad \a_4=-\a_+, \nonumber\\
\a_\pm&=&\sqrt{\delta\pm2d}, \quad
\delta=m^2+k^2+3\nu^2-a^2, \nonumber\\
d&=&\sqrt{(m^2+\nu^2)(k^2+2\nu^2-a^2)-(m\nu-ka)^2-\chi^2-c^2}.
\label{alfas2} \eea

\noindent The Ernst potentials defined by the axis data
(\ref{axis2}) were found to have the form \cite{EMR2}
\begin{eqnarray} \E=\frac{A-B}{A+B}, \quad
\Phi=\frac{C}{A+B}, \nonumber\\
\fl A=\{\delta[(m^2+\nu^2)^2+(m\nu-ka)^2]
-d^2(3m^2-k^2+\nu^2+a^2)\} \nonumber\\
\times(R_+-R_-)(r_+-r_-) -\a_+\a_-[\delta(m^2+\nu^2)-\chi^2-c^2]
\nonumber\\ \times(R_++R_-)(r_++r_-)
+2\a_+\a_-\{[(m^2+\nu^2)^2+(m\nu-ka)^2-d^2] \nonumber\\
\times(R_+R_-+r_+r_-)
+2\I d(m\nu-ka)(R_+R_--r_+r_-)\}, \nonumber\\
\fl B=4d\a_+\a_-(m+\I\nu)\{[m^2+\nu^2+\I(m\nu-ka)] \nonumber\\
\times(R_++R_--r_+-r_-)-d(R_++R_-+r_++r_-)\}, \nonumber\\
\fl C=4d(\chi+\I c)\{[m^2+\nu^2-\I(m\nu-ka)]
[\a_-(R_--R_+)-\a_+(r_--r_+)]
\nonumber\\
+d[\a_-(R_--R_+)+ \a_+(r_--r_+)]\}, \nonumber\\
R_\pm=\sqrt{\rho^2+(z\pm\a_+)^2}, \quad
r_\pm=\sqrt{\rho^2+(z\pm\a_-)^2}. \label{EF2}
\end{eqnarray}

\noindent The constants $m$, $a$, $\nu$, $k$ in
(\ref{axis2})--(\ref{EF2}) have the same meaning as in solution~I,
but $c$ and $\chi$ are the magnetic dipole and electric dipole
parameters, respectively.

The metric functions $f$, $\gamma$ and $\omega$ which appear in
the Papapetrou \cite{Pap} axisymmetric stationary line element
\begin{equation}\label{Papa}
\d s^2 = f^{-1}[e^{2\gamma} (\d\rho^2 + \d z^2) + \rho^2
\d\varphi^2]- f( \d t - \omega \d\varphi)^2
\end{equation} can be calculated with the aid of the general
formulae obtained in the paper \cite{RMM}: \bea &&f=\frac{E_+\bar
E_-+\bar E_+E_-+2F\bar F}{2E_-\bar E_-}, \quad
e^{2\gamma}=\frac{E_+\bar E_-+\bar E_+E_-+2F\bar F}{2K_0\bar K_0
r_1r_2r_3r_4}, \nonumber\\ &&\omega=\frac{2\,{\rm Im}(E_-\bar
H-\bar E_-G-F\bar I)} {E_+\bar E_-+\bar E_+E_-+2F\bar F},
\label{fgw_2} \eea

\noindent where the determinants $G$, $H$, $I$ and $K_0$ have the
form \bea \fl G=\left|\begin{array}{cccc} 0 & r_{1}+\a_{1}-z &
\ldots & r_{2N}+\a_{2N}-z \vspace{0.15cm}
\\ -1 & {\displaystyle \frac{r_1}{\a_1-\beta_1}} & \ldots &
{\displaystyle \frac{r_{4}}{\a_{4}-\beta_1}} \vspace{0.15cm}\\
-1 & {\displaystyle \frac{r_1}{\a_1-\beta_2}} & \ldots &
{\displaystyle \frac{r_{4}}{\a_{4}-\beta_2}} \vspace{0.25cm}\\ 0 &
{\displaystyle \frac{h_1(\a_1)}{\a_1-\bar\beta_1}} & \ldots &
{\displaystyle \frac{h_1(\a_{4})}{\a_{4}-\bar\beta_1}}
\vspace{0.15cm}\\ 0 & {\displaystyle
\frac{h_2(\a_1)}{\a_1-\bar\beta_2}} & \ldots & {\displaystyle
\frac{h_2(\a_{4})}{\a_{4}-\bar\beta_2}}\\
\end{array}\right|, \quad H=\left|\begin{array}{cccc} z & 1 & \ldots & 1
\vspace{0.15cm}\\ -\beta_1 & {\displaystyle
\frac{r_1}{\a_1-\beta_1}} & \ldots & {\displaystyle
\frac{r_{4}}{\a_{4}-\beta_1}} \vspace{0.15cm}\\ -\beta_2 &
{\displaystyle \frac{r_1}{\a_1-\beta_2}} & \ldots & {\displaystyle
\frac{r_{4}}{\a_{4}-\beta_2}} \vspace{0.25cm}\\ \bar e_1 &
{\displaystyle \frac{h_1(\a_1)}{\a_1-\bar\beta_1}} & \ldots &
{\displaystyle \frac{h_1(\a_{4})}{\a_{4}-\bar\beta_1}}
\vspace{0.15cm}\\ \bar e_2 & {\displaystyle
\frac{h_2(\a_1)}{\a_1-\bar\beta_2}} & \ldots & {\displaystyle
\frac{h_2(\a_{4})}{\a_{4}-\bar\beta_2}}\\
\end{array}\right|, \nonumber \\\nonumber\\
\fl I=\left|\begin{array}{ccccc} f_1+f_2 & 0 & f(\a_1) &
\ldots & f(\a_{4}) \vspace{0.15cm}\\ z & 1 & 1 & \ldots & 1 \\
-\beta_1 & -1 & {\displaystyle \frac{r_1}{\a_1-\beta_1}} & \ldots
& {\displaystyle \frac{r_{4}}{\a_{4}-\beta_1}} \vspace{0.15cm}\\
-\beta_2 & -1 & {\displaystyle \frac{r_1}{\a_1-\beta_2}} & \ldots
& {\displaystyle \frac{r_{4}}{\a_{4}-\beta_2}} \vspace{0.25cm}\\
\bar e_1 & 0 & {\displaystyle \frac{h_1(\a_1)}{\a_1-\bar\beta_1}}
& \ldots & {\displaystyle \frac{h_1(\a_{4})}{\a_{4}-\bar\beta_1}}
\vspace{0.15cm}\\ \bar e_2 & 0 & {\displaystyle
\frac{h_2(\a_1)}{\a_1-\bar\beta_2}} & \ldots &
{\displaystyle \frac{h_2(\a_{4})}{\a_{4}-\bar\beta_2}}\\
\end{array}\right|, \quad K_0=\left|\begin{array}{ccc} {\displaystyle
\frac{1}{\a_1-\beta_1}} & \ldots & {\displaystyle
\frac{1}{\a_{4}-\beta_1}} \vspace{0.15cm}\\
{\displaystyle \frac{1}{\a_1-\beta_2}} & \ldots & {\displaystyle
\frac{1}{\a_{4}-\beta_2}} \vspace{0.25cm}\\ {\displaystyle
\frac{h_1(\a_1)}{\a_1-\bar\beta_1}} & \ldots & {\displaystyle
\frac{h_1(\a_{4})}{\a_{4}-\bar\beta_1}} \vspace{0.15cm}\\
{\displaystyle \frac{h_2(\a_1)}{\a_1-\bar\beta_2}} & \ldots &
{\displaystyle
\frac{h_2(\a_{4})}{\a_{4}-\bar\beta_2}}\\
\end{array}\right|. \label{GHI_gen} \eea

It is clear that in all practical applications of the above
formulae for the metric functions, determinants (\ref{GHI_gen})
should be expanded and then evaluated for some particular axis
data with the aid of a computer program for analytical
calculations. In Appendix the reader can find the expansions of
the determinants (\ref{GHI_gen}) which have proved to be most
efficient for symbolic computer processing. The results of the
calculations of metric functions with the help of formulae (A.1),
(A.2) are given below. The functions $f$, $\gamma$ and $\omega$ of
both EMR solutions permit a unified representation, namely, \bea
f&=&\frac{A\bar A-B\bar B+C\bar C}{(A+B)(\bar A+\bar B)}, \quad
e^{2\gamma}=\frac{A\bar A-B\bar B+C\bar C}
{64\a_+^2\a_-^2 d^4 R_+R_-r_+r_-}, \nonumber\\
\omega&=&4\nu+\frac{{\rm Im}[\bar H(A+B)-G(\bar A+\bar B)-C\bar
I]} {A\bar A-B\bar B+C\bar C}, \label{fgw12} \eea where the
functions $A$, $B$, $C$, $R_\pm$, $r_\pm$ and quantities $\a_\pm$,
$d$ are defined by formulae (\ref{EF1}) and (\ref{alfas1}) in the
case of solution~I, and by formulae (\ref{EF2}) and (\ref{alfas2})
in the case of solution~II. The functions $G$, $H$ and $I$,
entering the expression of $\omega$, have the following
form\footnote{The formulae of this paper have been obtained and
checked using the Mathematica computer program \cite{Wol}.}: \bea
\fl G=-zB+2(m+\I\nu)A-2d\a_+\a_-(m^2+\nu^2-q^2-b^2)
\nonumber\\
\times[(\a_+-\a_-)(R_-r_--R_+r_+)+ (\a_++\a_-)(R_-r_+-R_+r_-)]
\nonumber\\
+4d\a_+\a_-[d-m^2-\nu^2+q^2+b^2+\I(k a-m\nu)] \nonumber\\
\times[\a_+(m+\I\nu)(R_--R_+)-2(m^2+\nu^2-q^2-b^2)
(R_-+R_+)]\nonumber\\
+4d\a_+\a_-[d+m^2+\nu^2-q^2-b^2-\I(k a-m\nu)] \nonumber\\
\times[\a_-(m+\I\nu)(r_--r_+)-2(m^2+\nu^2-q^2-b^2)
(r_-+r_+)], \nonumber\\
\fl H=z A-2(m+\I\nu)B +\a_+\a_-(m-\I\nu) \nonumber\\
\times\{(m^2+\nu^2-q^2-b^2) [(\a_+-\a_-)^2(R_+r_++R_-r_-)
\nonumber\\
+(\a_++\a_-)^2(R_+r_-+R_-r_+)]+2d(m+\I\nu) \nonumber\\
\times [(\a_+-\a_-)(R_+r_+-R_-r_-) +
(\a_++\a_-)(R_+r_--R_-r_+)] \nonumber\\
+ 4(R_+R_-+r_+r_-)[(m^2+\nu^2-q^2-b^2)(k^2-m^2+\nu^2-a^2)
\nonumber\\
-2(k a-m\nu)^2]+8\I d(k a-m\nu)(R_+R_--r_+r_-)\}
\nonumber\\
+4d\a_+\a_-(m+\I\nu)\{[m^2+\nu^2-q^2-b^2-\I(k a-m\nu)]
\nonumber\\
\times[\a_+(R_+-R_-)-\a_-(r_+-r_-)]
-d[\a_+(R_+-R_-)+\a_-(r_+-r_-)]\}, \nonumber\\
\fl I=2(q+\I b)(A+B)-[z+2(m+\I\nu)]C +2d\a_+\a_-(m-\I\nu)(q+\I b)
\nonumber\\ \times[(\a_+-\a_-) (R_+r_+-R_-r_-)
+(\a_++\a_-) (R_+r_--R_-r_+)] \nonumber\\
+4d\a_+\a_-(q+\I b) \{[m^2+\nu^2-q^2-b^2-\I(k a-m\nu)-d] \nonumber\\
\times[2(m-\I\nu)(R_++R_-)+\a_+(R_+-R_-)] -[m^2+\nu^2-q^2-b^2
\nonumber\\ -\I(k a-m\nu)+d]
[2(m-\I\nu)(r_++r_-)+\a_-(r_+-r_-)]\}, \label{GHI1} \eea in the
case of {\it solution~I}, and \bea \fl G=-zB+2(m+\I\nu)
A+2d\{(\a_+-\a_-) [(m^2+\nu^2)\a_+\a_-+\chi^2+c^2] (R_+r_+-R_-r_-)
\nonumber\\ +(\a_++\a_-)[(m^2+\nu^2)\a_+\a_--\chi^2-c^2]
(R_+r_--R_-r_+)\} \nonumber\\
+4d\a_+(m+\I\nu)\{[(k+\I a)^2-(m+\I\nu)^2] [m^2+\nu^2-d+\I(k
a-m\nu)] \nonumber\\ \times(r_+-r_-)
-2\a_-(m-\I\nu)[m^2+\nu^2+d-\I(k a-m\nu)](r_++r_-)\} \nonumber\\
+4d \a_-(m+\I\nu)\{[(k+\I a)^2-(m+\I\nu)^2] [m^2+\nu^2+d+\I(k
a-m\nu)] \nonumber\\ \times(R_--R_+)
+2\a_+(m-\I\nu)[m^2+\nu^2-d-\I(k a-m\nu)](R_-+R_+)\}, \nonumber\\
\fl H=z A-2(m+\I\nu)B+(m-\I\nu)(\a_+-\a_-)
\{(\a_+-\a_-)[(m^2+\nu^2)\a_+\a_-+\chi^2+c^2] \nonumber\\
\times (R_+R_-+r_+r_-) +2d\a_+\a_-(m+\I\nu)(R_+R_--r_+r_-)\}
\nonumber\\ +(m-\I\nu)(\a_++\a_-)
\{(\a_++\a_-)[(m^2+\nu^2)\a_+\a_--\chi^2-c^2] \nonumber\\
\times (R_+r_-+R_-r_+) +2d \a_+\a_-(m+\I\nu)(R_+r_--R_-r_+)\}
\nonumber\\ -4\a_+\a_-(m-\I\nu)\{[(m^2+\nu^2)^2+(k
a-m\nu)^2-d^2](R_+R_-+r_+r_-) \nonumber\\ -2\I d (k
a-m\nu)(R_+R_--r_+r_-)\}
+4d\a_+\a_-(m+\I\nu) \nonumber\\
\times \{[m^2+\nu^2-\I (k a-m\nu)] [\a_-(r_--r_+)-\a_+(R_--R_+)] \nonumber\\
+d[\a_-(r_--r_+)+\a_+(R_--R_+)]\},
\nonumber\\
\fl I=-(z+2m+2\I\nu)C+2(m-\I\nu)(\chi+\I c)\{[m^2+\nu^2+\I (k
a-m\nu)] [(\a_+\a_--\delta) \nonumber\\ \times (R_+r_++R_-r_-)
+(\a_+\a_-+\delta)(R_+r_-+R_-r_+) -2\a_+\a_- \nonumber\\
\times(R_+R_-+r_+r_-)] +2d^2(R_+-R_-)(r_+-r_-)
-2d\a_+\a_-(R_+R_--r_+r_-)\} \nonumber\\ +4d(\chi+\I c)
\{\a_+\a_-[3(m^2+\nu^2)-d+\I(k a-m\nu)](r_++r_-) \nonumber\\
+2\a_+(m-\I\nu)[m^2+\nu^2-d+\I(k a-m\nu)](r_+-r_-) -\a_+\a_-
\nonumber\\ \times [3(m^2+\nu^2) +d+\I(k a-m\nu)](R_++R_-)
-2\a_-(m-\I\nu) \nonumber\\ \times [m^2+\nu^2 +d+\I(k a-m\nu)]
(R_+-R_-)-8d\a_+\a_-(m+\I\nu)\} \label{GHI2} \eea in the case of
{\it solution~II}.

It is worthwhile mentioning that an arbitrary additive constant in
the expression of $\omega$ in (\ref{fgw12}) was chosen in such a
way that the constant $\omega_0$ in the definition of equatorially
antisymmetric spacetimes \cite{EMR2} were equal to zero, i.e.,
$\omega(\rho,z)=-\omega(\rho,-z)$ automatically.

\section{The two--body equilibrium problem in EMR spacetimes}

The expressions of the metric functions obtained in the previous
section can be used for the analysis of the equilibrium problem of
two counter--rotating constituents of the Kerr--Newman--NUT type.
In order these constituents to be in equilibrium, it is necessary
that the conditions \cite{Isr,Tom1,Tom2,DHo} \be \gamma=0, \quad
\omega=0 \label{cond_eq12} \ee

\begin{figure}[htb]
\centerline{\epsfysize=85mm\epsffile{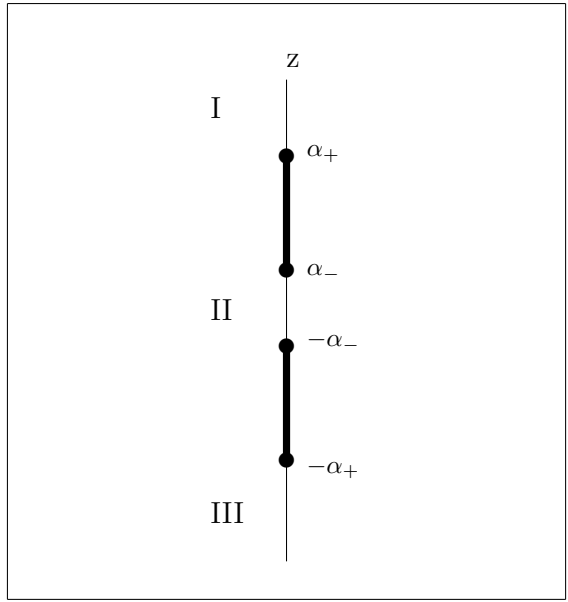}} \caption{Location
of sources in the solutions I and II.}
\end{figure}

\noindent are fulfilled on the part $\rho=0, \, |z|<\a_-$ of the
symmetry axis that separates the particles (region~II in Fig.~1).
An important advantage of the equatorially antisymmetric systems
is that the condition for $\omega$ is fulfilled automatically in
the region~II, what can be checked directly with the aid of
formulae (\ref{GHI1}) and (\ref{GHI2}), so that the condition for
the metric function $\gamma$ needs to be satisfied. The latter
condition leads to the algebraic equation \be \fl
(m^2+\nu^2-q^2-b^2)[(k^2+m^2+a^2+\nu^2)^2-4(k m+a\nu)^2
-\a_+\a_-\delta]=0, \label{bcg1} \ee in the case of {\it
solution~I}, and to the algebraic equations \be \fl
(\delta-\a_+\a_-)[(m^2+\nu^2)^2+(k
a-m\nu)^2]+d^2[\delta-4(m^2+\nu^2)+(1\pm 2)\a_+\a_-]=0,
\label{bcg2} \ee in the case of {\it solution~II}. It should be
mentioned that for both solutions $\gamma=0$ in the regions~I and
III, while $\omega=4\nu$ in the region~I and $\omega=-4\nu$ in the
region~III due to the presence of two semi--infinite NUT
singularities described in the paper \cite{MRu1}, the first
singularity extending from $\a_+$ to $+\infty$ and the second from
$-\a_+$ to $-\infty$. When $\nu=0$, both conditions
(\ref{cond_eq12}) are verified identically in the regions I and
III for solutions~I and II because the latter solutions become
asymptotically flat in that limit.

From (\ref{bcg1}) follows an important relation \be
m^2+\nu^2-q^2-b^2=0, \label{genPIW} \ee at which two
counter--rotating Kerr--Newman--NUT constituents are in
equilibrium independently of the distance between them. Formula
(\ref{genPIW}) generalizes the balance condition $m^2=q^2$ which
is verified by two Majumdar--Papapetrou equal charged masses
\cite{Maj,Pap2} (condition $m^2=q^2$ also determines two spinning
charged masses of the Perj\'es--Israel--Wilson type
\cite{Per,IWi,PRW}).

Furthermore, it can be shown that in the case of solution~I there
are no equilibrium states other than defined by the relation
(\ref{genPIW}) because the equation \be (k^2+m^2+a^2+\nu^2)^2 -4(k
m+a\nu)^2 -\a_+\a_-\delta=0 \label{c2_genBM} \ee has the roots
that cause $d=0$ in the denominator of the function $\gamma$ from
(\ref{GHI1}) and, therefore, should be discarded. Indeed, first
putting the term $\a_+\a_-\delta$ to the right--hand side of
equation (\ref{c2_genBM}) and then taking square of the resulting
equality, we arrive at the equation \bea \fl
[(k-m)^2+(a-\nu)^2][(k+m)^2+(a+\nu)^2] [(k^2+m^2+a^2+\nu^2)^2
\nonumber\\ -4(k m+a\nu)^2-(\delta+2b^2)^2
-4b^4+4b^2(\delta+2b^2)]=0. \eea

\noindent The first two factors of this equation become zero when
\be k=m, \quad a=\nu \qquad \hbox{and} \qquad k=-m, \quad a=-\nu,
\ee

\noindent and in both cases the solution~I reduces to the ordinary
NUT solution electrically and magnetically charged (see, e.g.,
Ref.~\cite{MMR}), that is, the two--body problem degenerates to
the case of a single body. Equating to zero the third factor and
solving the resulting equation with respect to $b^2$, we get \be
\fl b^2=\frac{1}{2}\Bigl(k^2+m^2+3\nu^2-a^2-2q^2
\pm\sqrt{(k^2+m^2+\nu^2+a^2)^2-4(k m+a\nu)^2}\Bigr). \label{b2}
\ee

\noindent Substituting now (\ref{b2}) in the expression of $d$
defined by (\ref{alfas1}) it is straightforward to check that
$d=0$ identically.

In the case of solution~II one can solve analytically the balance
equations (\ref{bcg2}) in exactly the same way as equation
(\ref{bcg1}) was solved for solution~I. For each choice of the
sign in (\ref{bcg2}) one then comes to an algebraic equation of
higher order than the initial one which, however, factorizes into
three factors, one of which is $d^2$, and the other two permit,
after equating them to zero, the solutions of the corresponding
equations with respect to $b^2$ similar to formula (\ref{b2}). The
direct substitution of the expressions for $b^2$ thus obtained
into equations (\ref{bcg2}) shows that they do not satisfy the
latter equations, thus being fictitious roots that must be
discarded. The case $d=0$ must be discarded too on the same
grounds as in the case of solution~I. Therefore, in the
equatorially antisymmetric systems of two magnetized masses there
are no equilibrium states under any choice of parameters, and
hence these masses are always supported by a massless strut
\cite{Isr} between them.

It should be emphasized that solutions~I and II are asymptotically
flat in the absence of the NUT parameter, in which case, by
construction, they are regular on the upper and lower parts of the
$z$--axis. In the presence of the NUT parameter, the conditions
$\gamma=0$, $\omega\ne 0$ are verified on those parts of the axis
for both solutions, giving rise to two semi--infinite massive
singularities of the NUT type \cite{MRu1}. We also mention that
although it is tempting, following paper \cite{LOl}, to redefine
the function $\gamma$, by adding a specific constant, in such a
way that the condition $\gamma=0$ is verified on the intermediate
part of the symmetry axis, this would only change the single
supporting intermediate strut $\gamma\ne 0$ of finite extension to
a pair of semi--infinite struts $\gamma\ne 0$ of Israel's type
(massless in the absence of the NUT parameter), thus only
worsening the situation and making the corresponding spacetimes
frankly unphysical.

\section{The multipole moments, basic limits, stationary limit
surfaces and ring singularities}

In the paper \cite{EMR2} the multipole structure of solutions~I
and II was not studied, so it would be of interest to clarify this
characteristic of the EMR solutions in some detail. We have
calculated the first four mass, angular momentum, electric and
magnetic multipole moments ($M_i$, $J_i$, $Q_i$ and $H_i$,
respectively) as these were defined by Simon \cite{Sim}. During
the calculations we have used the Hoenselaers--Perj\'es procedure
\cite{HPe} rectified by Sotiriou and Apostolatos \cite{SAp},
yielding the following expressions for the multipole moments: \bea
\fl M_0=2m, \quad M_2=2m(k^2-m^2+3\nu^2-a^2)-4ka\nu, \quad
M_1=M_3=0; \nonumber\\ \fl J_0=2\nu, \quad
J_2=4kma+2\nu(k^2-3m^2+\nu^2-a^2), \quad J_1=J_3=0; \nonumber\\
\fl Q_0=2q, \quad Q_2=2q(k^2-m^2+\nu^2-a^2)-4b(ka-m\nu), \quad
Q_1=Q_3=0; \nonumber\\ \fl H_0=2b, \quad
H_2=2b(k^2-m^2+\nu^2-a^2)+4q(ka-m\nu), \quad H_1=H_3=0
\label{mult1} \eea (the case of {\it solution~I}) and \bea \fl
M_0=2m, \quad
M_2=2m(k^2-m^2+3\nu^2-a^2)-4ka\nu, \quad M_1=M_3=0; \nonumber\\
\fl J_0=2\nu, \quad
J_2=4kma+2\nu(k^2-3m^2+\nu^2-a^2), \quad J_1=J_3=0; \nonumber\\
\fl Q_0=Q_2=0, \quad Q_1=2\chi, \quad
Q_3=2\chi(k^2-m^2+\nu^2-a^2)-4c(ka-m\nu); \nonumber\\ \fl
H_0=H_2=0, \quad H_1=2c, \quad
H_3=2c(k^2-m^2+\nu^2-a^2)+4\chi(ka-m\nu) \label{mult2} \eea (the
case of {\it solution~II}).

The above expressions support the physical meaning attributed to
the parameters of EMR solutions in the paper \cite{EMR2}. From
(\ref{mult1}) and (\ref{mult2}) follows that the main difference
between the two solutions lies in the structure of the
electromagnetic moments: in solution~I the odd moments $Q_{2n+1}$
and $H_{2n+1}$ are equal to zero, whereas in solution~II are equal
to zero the even moments $Q_{2n}$ and $H_{2n}$.

\begin{figure}[htb]
\centerline{\epsfysize=50mm\epsffile{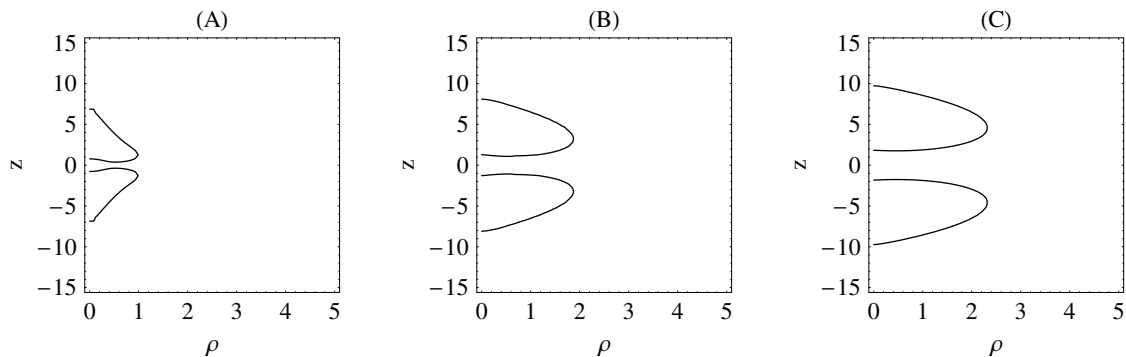}}
\caption{Particular SLS of solution~I demonstrating the {\it
growth} of SLS with increasing NUT parameter $\nu$. In all three
cases $m=4$, $k=3$, $a=1$, $q=b=1/2$, but $\nu$ varies: (A)
$\nu=1$, (B) $\nu=2$, (C) $\nu=4$.}
\end{figure}

In the absence of the electromagnetic field, both EMR solutions
reduce to the same special vacuum spacetime for two
counter--rotating Kerr--NUT masses, and this limit belongs to the
double--Kerr family of solutions of Kramer and Neugebauer
\cite{KNe}. It is interesting that by further setting $k=m$,
$a=\nu$, one arrives at the single NUT solution \cite{NTU} with
the total mass $2m$ and the NUT parameter $2\nu$ which, as was
demonstrated by Manko and Ruiz \cite{MRu1}, shares the property of
being equatorially antisymmetric. Solution~I reduces to the
Bret\'on--Manko electrovac solution \cite{BMa} for two
counter--rotating Kerr--Newman masses when the NUT parameter $\nu$
and the magnetic charge $b$ are equal to zero. The physically most
interesting subclass of solution~II is defined by vanishing
parameters $\nu$ and $\chi$, in which case this EMR solution
represents two counter--rotating magnetized masses, giving the
first example of solutions of this kind known in the literature.

\begin{figure}[htb]
\centerline{\epsfysize=50mm\epsffile{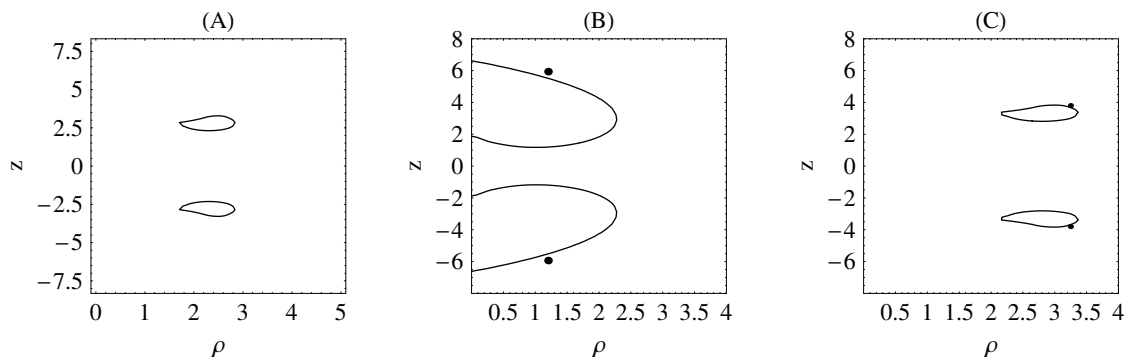}} \caption{Specific
SLS of solution~I: (A) The case of two {\it hyperextreme}
constituents defined by $m=2$, $k=a=3$, $\nu=q=b=1$; (B) Two
subextreme constituents with {\it negative} masses ($m=-4$, $k=3$,
$a=1$, $\nu=q=b=1/2$) and two ring singularities located at
$\rho\simeq 1.2$, $z\simeq\pm 5.935$; (C) Two hyperextreme
constituents with {\it negative} masses ($m=-2$, $k=a=3$,
$\nu=q=b=1/2$) and two ring singularities located at $\rho\simeq
3.246$, $z\simeq\pm 3.801$. The dots in (B) and (C) denote ring
singularities.}
\end{figure}

The stationary limit surfaces (SLS) which are defined by the
equation $f=0$, in the case of EMR spacetimes display several
interesting properties worthy of mentioning here. In figures~2 and
3 we have plotted several particular SLS of solution~I, while in
figures~4 and 5 one can find particular SLS of solution~II.
Figures~2 and 4 clearly demonstrate a completely different
evolution of SLS in solutions I and II as function of the NUT
parameter: in solution~I (see figure~2) the SLS grows with
increasing $\nu$, while in solution~II (figure~4) the increase of
$\nu$ causes the degeneration of SLS. In figures~3(A) and 5(A) the
typical SLS of toroidal type show a clear similarity in the case
of two hyperextreme constituents with {\it positive} masses for
both solutions~I and II. Nonetheless, in the case of constituents
with {\it negative} masses the corresponding ring singularities
develop differently. Indeed, it follows from figure~3(B,C) that
ring singularities accompanying the negative mass in solution~I
are located {\it outside} the SLS, though very close to it. At the
same time, looking at figure~5(B,C), one can see that the ring
singularities of solution~II, either in the subextreme or in the
hyperextreme cases, lie on the SLS, exactly as in the pure vacuum
case. We remind that ring singularities arise as solutions of the
equation $A+B=0$.

\begin{figure}[htb]
\centerline{\epsfysize=50mm\epsffile{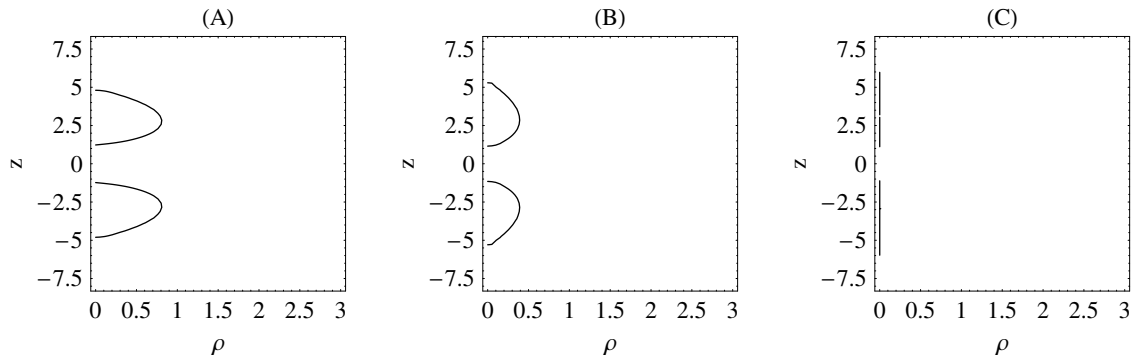}}
\caption{Particular SLS of solution~II demonstrating the {\it
degeneration} of SLS with growing NUT parameter $\nu$. In all
three cases $m=2$, $k=3$, $a=1$, $c=\chi=1/2$, while $\nu$ varies
in the following way: (A) $\nu=1/2$, (B) $\nu=1$, (C) $\nu=3/2$.}
\end{figure}

\begin{figure}[htb]
\centerline{\epsfysize=50mm\epsffile{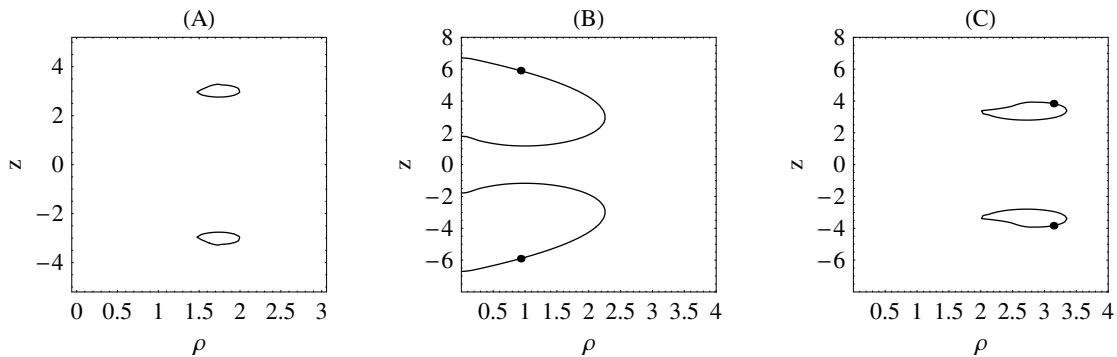}} \caption{Specific
SLS of solution~II: (A) The case of two {\it hyperextreme}
constituents defined by $m=1$, $k=3$, $a=2$, $\nu=1/4$,
$c=\chi=1/2$; (B) Two subextreme constituents with {\it negative}
masses ($m=-4$, $k=3$, $a=1$, $\nu=c=\chi=1/2$) and two ring
singularities located at $\rho\simeq 0.932$, $z\simeq\pm 5.908$;
(C) Two hyperextreme constituents with {\it negative} masses
($m=-2$, $k=a=3$, $\nu=c=\chi=1/2$) and two ring singularities
located at $\rho\simeq 3.142$, $z\simeq\pm 3.836$. The dots in (B)
and (C) denote ring singularities.}
\end{figure}

\section{Conclusions}

The new symmetry discovered and described by Ernst {\it et al}
\cite{EMR1} permits a systematic study of a large class of
counter--rotating masses within the framework of general
relativity. Thanks to the paper \cite{EMR1}, and partly to the
paper \cite{MRu1}, we know for instance that the well--known NUT
solution belongs to the family of equatorially antisymmetric
spacetimes; it is also clear now that the word `antisymmetric'
applied more than a decade ago by Bret\'on and Manko to a system
of two equal counter--rotating Kerr--Newman particles in
\cite{BMa} was quite appropriate. While the NUT solution
represents a single body accompanied by two semi--infinite
singularities, the new EMR 6--parameter solutions already describe
the two--body systems endowed with singularities of the NUT type.
In the present paper we have constructed all metrical fields for
both EMR solutions and have solved analytically the associated
equilibrium problems, obtaining the genuine equilibrium states
only for a particular subclass of solution~I defined by the
relation $m^2+\nu^2=q^2+b^2$ (the two counter-rotating
constituents become then hyperextreme). Some physical properties
of the EMR spacetimes have been also studied. As a final remark we
would like to observe that in view of the physical differences
existing in the equatorially symmetric case between the systems
with even and odd number of particles (see, e.g.,
\cite{IKh,MRM,MMRu}) it may be interesting to consider the
three-body equatorially antisymmetric solutions and compare them
with the EMR two--body spacetimes.

\ack

We thank Professor Vladimir S. Manko for helpful discussions. JSH
also thanks Erasmo G\'omez for some technical computer assistance.
This work was partially supported by Project 45946--F from CONACyT
of Mexico.

\appendix
\section*{Appendix. Expansions of the determinants
$E_\pm$, $F$, $G$, $H$, $I$, $K_0$}

\setcounter{section}{1}

Since the determinants (\ref{EF_gen}), (\ref{GHI_gen}) contain the
coordinates $\rho$ and $z$ only through functions $r_n$, it is
advantageous to expand these determinants over the two lines in
which $r_n$ appear, using the Laplace rule. The resulting
expressions for the determinants employed in the analytical
computer codes for obtaining formulae (\ref{GHI1}), (\ref{GHI2})
have the form \bea \fl E_\pm= \Lambda\pm\Gamma=A\mp B, \nonumber\\
\fl \Lambda= \sum\limits_{1\le i<j\le 4} (-1)^{i+j} r_i
r_j(\a_i-\a_j)R_k R_l\tilde R_i\tilde R_j \nonumber\\ \times
[(\a_k-\bar\beta_2)(\a_l-\bar\beta_1)h_1(\a_k)h_2(\a_l)
-(\a_k-\bar\beta_1)(\a_l-\bar\beta_2)h_1(\a_l)h_2(\a_k)],
\nonumber\\ (k<l; \quad k,l\ne i,j) \nonumber\\ \fl \Gamma=
\sum\limits_{i=1}^4 (-1)^{i} r_i R_j R_k R_l\tilde R_i \{\tilde
R_j \nonumber\\
\times [(\a_k-\bar\beta_2)(\a_l-\bar\beta_1)h_1(\a_k)h_2(\a_l)
-(\a_k-\bar\beta_1)(\a_l-\bar\beta_2)h_1(\a_l)h_2(\a_k)]
\nonumber\\
-\tilde R_k[(\a_j-\bar\beta_2)
(\a_l-\bar\beta_1)h_1(\a_j)h_2(\a_l)
-(\a_j-\bar\beta_1)(\a_l-\bar\beta_2)h_1(\a_l)h_2(\a_j)]
\nonumber\\ + \tilde
R_l[(\a_j-\bar\beta_2)(\a_k-\bar\beta_1)h_1(\a_j)h_2(\a_k)
-(\a_j-\bar\beta_1)(\a_k-\bar\beta_2)h_1(\a_k)h_2(\a_j)]\},
\nonumber\\ \fl F=C= \sum\limits_{i=1}^4 (-1)^{i+1} r_i R_j R_k
R_l\tilde R_i\{\tilde R_j f(\a_j) \nonumber\\
\times[(\a_k-\bar\beta_2)(\a_l-\bar\beta_1)h_1(\a_k)h_2(\a_l)
-(\a_k-\bar\beta_1)(\a_l-\bar\beta_2)h_1(\a_l)h_2(\a_k)]
\nonumber\\ - \tilde R_k
f(\a_k)[(\a_j-\bar\beta_2)(\a_l-\bar\beta_1)h_1(\a_j)h_2(\a_l)
\nonumber\\
-(\a_j-\bar\beta_1)(\a_l-\bar\beta_2)h_1(\a_l)h_2(\a_j)]
\nonumber\\ + \tilde R_l
f(\a_l)[(\a_j-\bar\beta_2)(\a_k-\bar\beta_1)h_1(\a_j)h_2(\a_k)
\nonumber\\
-(\a_j-\bar\beta_1)(\a_k-\bar\beta_2)h_1(\a_k)h_2(\a_j)]\},
\nonumber\\
(j<k<l; \quad j,k,l\ne i) \nonumber\\
\fl G=z\Gamma-(\beta_1+\beta_2)\Lambda + \sum\limits_{1\le i<j\le
4}(-1)^{i+j}r_i r_j (\a_i^2-\a_j^2)R_k R_l \tilde R_i\tilde R_j
\nonumber\\ \times
[(\a_k-\bar\beta_2)(\a_l-\bar\beta_1)h_1(\a_k)h_2(\a_l)
-(\a_k-\bar\beta_1)(\a_l-\bar\beta_2)h_1(\a_l)h_2(\a_k)]
\nonumber\\ + \sum\limits_{i=1}^4 (-1)^{i+1} r_i R_j R_k R_l\tilde
R_i \{\a_j\tilde R_j \nonumber\\ \times
[(\a_k-\bar\beta_2)(\a_l-\bar\beta_1)h_1(\a_k)h_2(\a_l)
-(\a_k-\bar\beta_1)(\a_l-\bar\beta_2)h_1(\a_l)h_2(\a_k)]
\nonumber\\ - \a_k\tilde
R_k[(\a_j-\bar\beta_2)(\a_l-\bar\beta_1)h_1(\a_j)h_2(\a_l)
\nonumber\\
-(\a_j-\bar\beta_1)(\a_l-\bar\beta_2)h_1(\a_l)h_2(\a_j)]
\nonumber\\ + \a_l\tilde
R_l[(\a_j-\bar\beta_2)(\a_k-\bar\beta_1)h_1(\a_j)h_2(\a_k)
\nonumber\\
-(\a_j-\bar\beta_1)(\a_k-\bar\beta_2)h_1(\a_k)h_2(\a_j)]\},
\nonumber\\ \fl H=z\Lambda-(\beta_1+\beta_2)\Gamma +
\sum\limits_{i=1}^4 (-1)^{i} r_i \a_i R_j R_k R_l\tilde R_i
\nonumber\\
\times\{\tilde
R_j[(\a_k-\bar\beta_2)(\a_l-\bar\beta_1)h_1(\a_k)h_2(\a_l)
-(\a_k-\bar\beta_1)(\a_l-\bar\beta_2)h_1(\a_l)h_2(\a_k)]
\nonumber\\ - \tilde
R_k[(\a_j-\bar\beta_2)(\a_l-\bar\beta_1)h_1(\a_j)h_2(\a_l)
-(\a_j-\bar\beta_1)(\a_l-\bar\beta_2)h_1(\a_l)h_2(\a_j)]
\nonumber\\ + \tilde
R_l[(\a_j-\bar\beta_2)(\a_k-\bar\beta_1)h_1(\a_j)h_2(\a_k)
-(\a_j-\bar\beta_1)(\a_k-\bar\beta_2)h_1(\a_k)h_2(\a_j)]\}
\nonumber\\ - \sum\limits_{1\le i<j\le 4} (-1)^{i+j} r_i
r_j(\a_i-\a_j)R_k R_l\tilde R_i\tilde R_j \nonumber\\ \times \{
\tilde R_k[\bar e_1(\a_l-\bar\beta_1)h_2(\a_l) -\bar
e_2(\a_l-\bar\beta_2)h_1(\a_l)] \nonumber\\ -\tilde R_l[\bar
e_1(\a_k-\bar\beta_1)h_2(\a_k) -\bar
e_2(\a_k-\bar\beta_2)h_1(\a_k)]\}, \nonumber\\
\fl I=(f_1+f_2)(\Lambda-\Gamma)+(\beta_1+\beta_2-z)F \nonumber\\
+ \sum\limits_{1\le i<j\le 4} (-1)^{i+j} r_i r_j(\a_i-\a_j)R_k
R_l\tilde R_i\tilde R_j \nonumber\\ \times \{\bar
e_2[(\a_l-\bar\beta_2)h_1(\a_l)\tilde R_k f(\a_k)
-(\a_k-\bar\beta_2)h_1(\a_k)\tilde R_l f(\a_l))] \nonumber\\
-\bar e_1[(\a_l-\bar\beta_1)h_2(\a_l)\tilde R_k f(\a_k)
-(\a_k-\bar\beta_1)h_2(\a_k)\tilde R_l f(\a_l))]\} \nonumber\\
+ \sum\limits_{i=1}^4 (-1)^{i+1} r_i R_j R_k R_l\tilde R_i
\nonumber\\ \times\{ [\bar e_1(\a_j-\bar\beta_1)h_2(\a_j)-\bar
e_2(\a_j-\bar\beta_2)h_1(\a_j)] [ f(\a_k)-f(\a_l)]\tilde R_k\tilde
R_l \nonumber\\ -[\bar e_1(\a_k-\bar\beta_1)h_2(\a_k)-\bar
e_2(\a_k-\bar\beta_2)h_1(\a_k)] [ f(\a_j)-f(\a_l)]\tilde R_j\tilde
R_l \nonumber\\ +[\bar e_1(\a_l-\bar\beta_1)h_2(\a_l)-\bar
e_2(\a_l-\bar\beta_2)h_1(\a_l)] [
f(\a_j)-f(\a_k)]\tilde R_j\tilde R_k\} \nonumber\\
+ \sum\limits_{i=1}^4 (-1)^{i} r_i \a_i R_j R_k R_l\tilde R_i
\{\tilde R_j f(\a_j) \nonumber\\
\times [(\a_k-\bar\beta_2)(\a_l-\bar\beta_1)h_1(\a_k)h_2(\a_l)
-(\a_k-\bar\beta_1)(\a_l-\bar\beta_2)h_1(\a_l)h_2(\a_k)]
\nonumber\\ - \tilde R_k
f(\a_k)[(\a_j-\bar\beta_2)(\a_l-\bar\beta_1)h_1(\a_j)h_2(\a_l)
\nonumber\\
-(\a_j-\bar\beta_1)(\a_l-\bar\beta_2)h_1(\a_l)h_2(\a_j)]
\nonumber\\ + \tilde R_l
f(\a_l)[(\a_j-\bar\beta_2)(\a_k-\bar\beta_1)h_1(\a_j)h_2(\a_k)
\nonumber\\
-(\a_j-\bar\beta_1)(\a_k-\bar\beta_2)h_1(\a_k)h_2(\a_j)]\}
\nonumber\\ - R_1 R_2 R_3 R_4 \sum\limits_{1\le i<j\le
4}(-1)^{i+j}\tilde
R_i\tilde R_j[ f(\a_i)-f(\a_j)] \nonumber\\
\times[(\a_k-\bar\beta_2)(\a_l-\bar\beta_1)h_1(\a_k)h_2(\a_l)
-(\a_k-\bar\beta_1)(\a_l-\bar\beta_2)h_1(\a_l)h_2(\a_k)],
\nonumber\\ \fl K_0= \sum\limits_{1\le i<j\le 4} (-1)^{i+j}
(\a_i-\a_j)R_k R_l\tilde R_i\tilde R_j
[(\a_k-\bar\beta_2)(\a_l-\bar\beta_1)h_1(\a_k)h_2(\a_l) \nonumber\\
-(\a_k-\bar\beta_1)(\a_l-\bar\beta_2)h_1(\a_l)h_2(\a_k)], \eea
where \be R_k=(\a_k-\beta_1)(\a_k-\beta_2), \quad \tilde
R_k=(\a_k-\bar\beta_1)(\a_k-\bar\beta_2) \ee are the constant
objects first introduced by Manko and Ruiz for the vacuum soliton
solution \cite{MRu2}. Note that the expressions (A.1) are given
without the common factor $(\beta_1-\beta_2)/(\prod_{n=1}^4
R_n\tilde R_n)$.

\end{document}